\documentclass[fleqn,10pt]{wlscirep}
\usepackage[utf8]{inputenc}
\usepackage[T1]{fontenc}
 \usepackage{float} 
 \usepackage{cite}
\usepackage{graphicx}
\usepackage{eucal}
\usepackage{amsmath, tabularx}
\usepackage{bigstrut}
\usepackage[font=small,labelfont=bf,tableposition=top]{caption}
\usepackage{nccmath}
\usepackage[sort&compress, numbers]{natbib}
\usepackage{dirtytalk}
\usepackage{multirow}

\title{Evaluating clinical diversity and plausibility of synthetic capsule endoscopic images}
\author[1,*]{Anuja Vats}
\author[1]{Marius Pedersen}
\author[1,2]{Ahmed Mohammed}
\author[1,3]{Øistein Hovde}

\affil[1]{Department of Computer Science, NTNU, 2819, Gjøvik, Norway}
\affil[2]{SINTEF Digital, Smart Sensor Systems, Oslo, Norway}
\affil[3]{Innlandet Hospital Trust, 2819, Gjøvik, Norway}

\affil[*]{anuja.vats@ntnu.no}


\begin{abstract}
Wireless Capsule Endoscopy (WCE) is being increasingly used as an alternative imaging modality for complete and non-invasive screening of the gastrointestinal tract. Although this is advantageous in reducing unnecessary hospital admissions, it also demands that a WCE diagnostic protocol be in place so larger populations can be effectively screened. This calls for training and education protocols attuned specifically to this modality. Like training in other modalities such as traditional endoscopy, CT, MRI, etc., a WCE training protocol would require an atlas comprising of a large corpora of images that show vivid descriptions of pathologies and abnormalities, ideally observed over a period of time. Since such comprehensive atlases are presently lacking in WCE, in this work, we propose a deep learning method for utilizing already available studies across different institutions for the creation of a realistic WCE atlas using StyleGAN. We identify clinically relevant attributes in WCE such that synthetic images can be generated with selected attributes on cue. Beyond this, we also simulate several disease progression scenarios. The generated images are evaluated for realism and plausibility through three subjective online experiments with the participation of eight gastroenterology experts from three geographical locations and a variety of years of experience. The results from the experiments indicate that the images are highly realistic and the disease scenarios plausible. The images comprising the atlas are available publicly for use in training applications as well as supplementing real datasets for deep learning.

\end{abstract}
\begin{document}

\maketitle
%
%
\thispagestyle{empty}

\section*{Introduction}
\label{sec:intro}
Wireless Capsule Endoscopy (WCE) has led to major advances in the diagnosis and treatment of small bowel diseases. As it is non-invasive and a comparatively more comfortable procedure than colonoscopy, it is also preferred by many patients for investigation of the colon \citep{romero2014CCEVSWCE}. The limited capacity for patient examinations during the COVID 19 pandemic \citep{halder2022colonCCEVSWCE} further propelled its use as it is less resource and time consuming. WCE has been established as a promising alternative for the diagnosis and monitoring of various diseases, including lesions such as polyps, tumors \citep{eliakim2009prospective,van2009capsule}, Ulcerative Colitis (UC) \citep{shi2017prospectiveCCEVSWCE}, Chron's disease \citep{goran2018capsulechron} as well as other markers related to inflammatory bowel diseases such as inflammatory grading (occurrence of bleeding, ulcers, erosions), mucosal healing (edema, changes in mucosal vascularity), etc.\citep{holleran2014colon,gay2010could}. 

Reading through a WCE video is often a tedious and demanding endeavor for clinicians. Videos are not only long (between 8 to 11 hours in most cases) but require undivided attention throughout to avoid missing smaller pathologies, such as polyps. Additionally, clinicians must reliably identify gastrointestinal landmarks for near-accurate positioning of abnormalities for reporting and subsequent monitoring. This expertise requires training for traditional endoscopists, even more so considering the diagnostically relevant differences that arise from the two modalities. Differences such as the direction of traversal (WCE traversing opposite to colonoscopy), organ appearance and colors (due to vast lighting differences) and inability for improving visibility through air-insufflation and water cleansing (as in endoscopy) tend to affect how diagnoses are carried out. In traditional endoscopy, medical trainees typically see thousands of images of different pathologies under progression or regression of the disease to identify disease markers. Despite training according to well-established diagnostic protocols, current studies indicate possible gaps in gastroenterology training \citep{maida2020current}, pointing to an even greater inadequacy in training for WCE diagnosis. Therefore, a need for additional training for competency in WCE \citep{rajan2013training,koffas2019training}.

One hindrance to such WCE based training protocol is the lack of a diverse and more complete atlas describing visual symptoms of diseases as seen in WCE. Many small scale atlases exist, including those from capsule manufacturers (e.g, Medtronic\texttrademark), however they fall exceedingly short in descriptions of diseases often with just one or few examples of a certain illness. Remedying this requires larger atlases with diverse medical conditions and severity descriptions, which, in turn require diverse patient populations and subsequent data labeling. Furthermore, ideal severity level descriptions require monitoring of patients with a specific disease over possibly long follow-up periods. The high diagnostic times per patient of WCE currently make the creation of such an atlas very difficult. In this work, we propose a deep learning method for the creation and quality evaluation of precisely such an atlas for WCE from previously collected (unlabeled) data for medical training and education.

The prediction of the prognosis of the disease and subsequent treatment is critically dependent on identifying the peculiarities associated with different severities. In lieu of real images, synthetic images could be used to simulate these peculiarities for various disease conditions. Naturally, considering the critical nature of this application, it is impending that (a.) synthetic images be consistent in visual symptoms in addition to being realistic, meaning disease markers can be learned independently from location, illuminations, intestinal fluid, etc., and (b.) images can be generated with desired pathology and severity on cue. 
To this end, we use existing available studies stored with two medical institutions for the creation of new and diverse WCE images/scenarios on cue. Furthermore, we take advantage of the diversity in pathologies between patients to learn progressive/regressive disease patterns and use them to simulate new and realistic disease scenarios for clinical training and education purposes, as well as means to compensate for real datasets. We use the style-based generative adverserial network StyleGAN2\cite{Karras2019stylegan2} to create synthetic images completely unsupervised and subsequently mine clinically relevant attributes for the creation of an atlas. We evaluate the plausibility and usability of our synthetic data through a series of subjective online experiments (Sec.\ref{sec3}) performed by eight gastrointestinal experts (seven gastroenterologists and one gastrosurgeon) from three different geographical locations and with experience ranging from 3 to 36 years. To the best of our knowledge, this is the first study to realistically simulate and evaluate realism in WCE images and disease progression scenarios. The main contributions of this work are as follows:
\begin{itemize}
\item Creation of a realistic WCE atlas/dataset with diverse clinical scenarios (with and without pathologies).
    \item Evaluation of clinical-realism of synthetically generated high-resolution WCE images through subjective experiments.
    \item Realistic simulation of clinically consistent disease progressions (without access to real progressions) and its plausibility evaluation.
\end{itemize}

\section{Related Work}\label{sec2}
Deep learning-based solutions for computer-aided diagnosis and detection of diseases have inevitably led to the need for high-quality medical data sets in different modalities \cite{koshino2021narrative} in addition to real data. The primary purpose of augmentation with synthetic data has been to train deeper and more data-hungry models for better performance \cite{kazuhiro2018generative, carver2021improvement,turan2021effectiveness,diamantis2022endovae,vats2021EUVIP}. In WCE, in addition to video enhancement \cite{mohammed2018deep}, supervised and semi-supervised abnormality classification \cite{vats2021learning, vats2022multichannel} and detection \cite{gilabert2022artificial}, synthetic data generation has become increasingly of interest \cite{turan2021effectiveness, diamantis2022endovae}. This is due to a pervasive data adversity compounded with the need for computer intervention to help experts in WCE. Traditional methods of synthetic image creation depend heavily on priors such as anatomy, colors, etc. \cite{turan2021effectiveness}, and do not scale easily when diverse pathological descriptions are taken into account. Recent deep learning-based generation methods in WCE although reasonably realistic \cite{diamantis2022endovae, xiao2022wce}, have little or no control over the clinical attributes comprising each generated image (hereby referred to as generates). Methods that enable the control of meaningful clinical markers of disease so that realistic images with desirable attributes can be simulated on cue is a fairly new research direction \cite{kim2022ganeval, segal2021evaluating, fetty2020latent, shen2020interpreting} that this work delves into. Moreover, the evaluation of synthetic datasets using pretrained object detectors as in \cite{xiao2022wce} while useful in assessing realness to some extent, does not reliably measure the diversity within the dataset. We believe, medical experts play an irreplaceable role in the evaluation of synthetic medical data through closely scrutinizing all aspects of generated images. \\
Generative Adverserial Networks (GANs) \cite{NIPS2014_gans} have single handedly transformed image synthesis in a variety of fields \cite{radford2015unsupervised, karras2019style, costa2017towards}. GANs typically learn the data distributions by constantly generating and differentiating increasingly complex generates from reals. Consequently, completely new samples can be generated from the learned distribution after training. The introduction of a style-based architecture \cite{Karras2019stylegan2, kim2022ganeval} significantly improved image synthesis, by allowing a higher degree of control over the generation process. In WCE, this translates to identifying clinically relevant attributes/biomarkers in the latent space of the model \cite{shen2020interpreting,fetty2020latent} and using them as inputs for data generation. Such controlled generation has been explored in modalities like retinal, CT, X-ray, etc. \cite{skandarani2021ganseval, segal2021evaluating,kim2022ganeval}. However, the aforementioned medical domains are simpler than WCE. Medical domains such as retinal and X-ray imaging exhibit familiar components, that are repetitive. Furthermore, most studies focus on one of the few particular pathology types against this familiar background during synthetic generation. We propose to expand these boundaries and test synthetic data generation on gastrointestinal images with varied pathological (polyps, ulcer, inflammation, vascular pattern , edema, fibrin, etc.), anatomical (ascending, descending, transverse, small bowel), orientation (rotation within lumen, muscosal facing, lumen facing), gastroninetstinal content (debris, bubbles), and capsule modality (pillcam vs. other) variations.
In addition to examining the latent space for realistic generation of individual data points, such as images, we also focus on identifying clinical biomarkers that are reliable indicators of disease and use those to generate realistic disease progressions. Additionally, we perform a critical evaluation of the generated data beyond the commonly employed Turing tests \cite{segal2021evaluating, kim2022ganeval, skandarani2021ganseval} to concretely measure the plausibility and efficacy of synthetic data in WCE.

\section{Methodology.}\label{sec3}
\begin{figure}[!htbp]
\centering
\includegraphics[width=6in, height=8in]{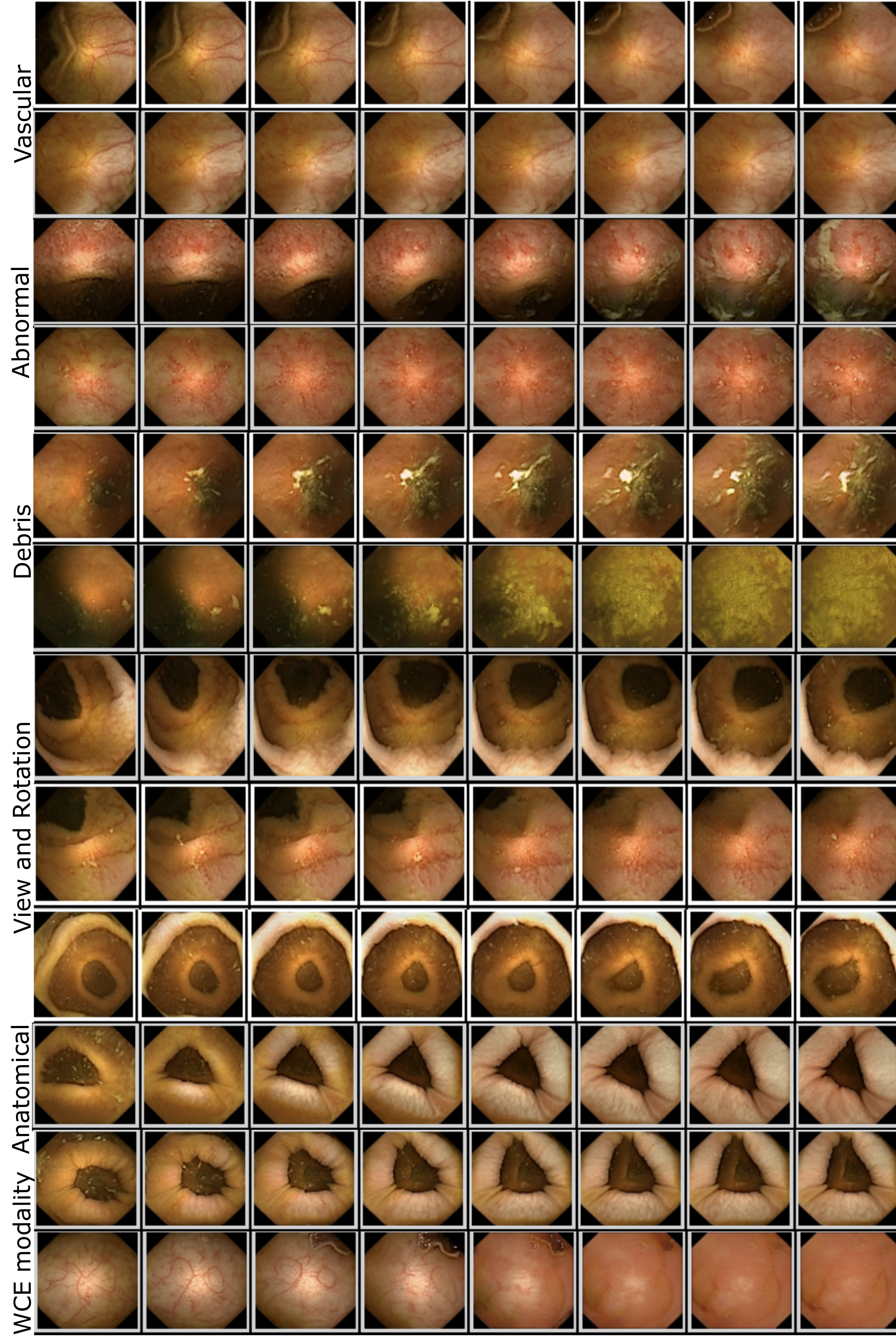}
\caption{Figure illustrates synthetic images from the atlas with the following variations- a.) Vascular: Variations in the vascular pattern underneath the mucosa, used as an indicator of tissue health b.) Abnormal: variations that are pathological in nature such as development of inflammation, ulcer, edema etc. c.) Debris: variations that simulate various levels and types of occlusions expected in WCE images d.) View and Rotation: Variations simulating different view points as well as free capsule rotation as it traverses e.) Anatomical : variations relating to peristalsis as well as those arising from different parts of the tract f.) WCE modality: variations that reflect changes from one capsule modality to another (such as change in organ colors, illumination etc.) }
\label{ATLAS}
\end{figure}
In this work, we utilize the generative power of StyleGAN2 \citep{Karras2019stylegan2, karras2019style} for unsupervised image generation at high resolution (512x512). A random variable $\textbf{z}\in \mathcal{Z}$ is input to a dense mapping network $f$ that transforms the incoming latent variables into d-dimensional vectors $\mathcal{\textbf{w}}$ lying in the intermediate latent space $\mathcal{W}$. The learned affine transformations $\textbf{A}\in\mathbb{R}^{mxd}$ transform a latent code $\textbf{w}$ into a style vector $\textbf{y}$ as shown in Eq.\ref{eq:eq1}, where $\textbf{b} \in \mathbb{R}^{m}$ is the bias. The generator $g$ uses a constant value vector $\textbf{k}$ as an initial point for image generation that is affected by the style vector $\textbf{y}$ (via Adaptive Instance Normalization (AdaIN)) and stochastic noise injection at each layer. Learning through $f$ in this way allows the intermediate space $\mathcal{W}$ to exhibit a degree of disentanglement such that the attributes underlying the data lie along mutually orthogonal directions. Generating along these attributes results in conditional manipulation of generated images \citep{fetty2020latent,vatsicassp}. 
\begin{equation}
    \centering
    \textbf{y} = \textbf{Aw}+\textbf{b}
    \label{eq:eq1}
\end{equation}

\noindent\textbf{Attribute discovery in $\mathcal{W}$:} The latent space $\mathcal{W}$ of a trained model can be interpreted as a high dimensional space consisting of factors that underlie the dataset. For WCE these factors are mainly related to anatomical, pathological, illumination, capsule modality, and gastrointestinal content (refer to Fig.\ref{ATLAS}). The first step into controlled image generation is to identify significant factors/attributes and consequently label them. In the natural image domain, this is achieved through pre-trained attribute detectors which are largely unavailable in medical domains. We perform closed-form SEmantic FActorization (SeFA \cite{shen2021closed}) on the affine transformation matrix $\textbf{A}$ in Eq.\ref{eq:eq1} such that the first $j$ eigen vectors of $\textbf{A}^T\textbf{A}$ correspond to the largest variations in the data (seen in Fig.\ref{ATLAS}). To label the variations, we performed t-distributed Stochastic Neighbor Embedding (t-SNE) \cite{van2008visualizing} clustering in image space (since all variations are orthogonal in latent space and measures of similarity do not apply) by seeding the space with disease-prototype images. These prototypical images are generated images selected by a gastrointestinal expert as being both realistic and prototypical to abnormalities (such as ulcer, inflammation, polyp, etc.). All attributes along which one or more images lie in the neighborhood of prototypical images are selected as pathology relevant, while others contain largely non-pathological variations (such as debris, modality, etc.).

\noindent\textbf{Image generation along attribute $a$:} To create $n$ images $\{i_1, i_2, ... i_n\}$ along $a\in\mathbb{R}^{512}$, we first create $n$ latent codes along $a$ in $\mathcal{W}$ given by $\{ w_1, w_2, ... w_n\}$ such that $w_i = w_n - \alpha_j *a$ and $\alpha $ varies linearly in interval $[A,B]$. $A$ and $B$ are chosen for each attribute such that largest variations can be simulated between them. Intervals between 0 and 50 were observed to be sufficient for most attributes. The parameter $\alpha$ also plays a key role, as a smaller step size results in images with very minute differences (as seen in Fig.\ref{ATLAS}) whereas a larger $\alpha$ can result in missing out of images with subtle abnormalities. Although $\alpha$ is a hyperparameter and depends largely on the modality, for this work $\alpha=2$ was found to be suitable for most variations.

The styleGAN2 model was trained on approximately 200k images (dataset detail in Sec.\ref{sec:analyses}) without progressive growing on a Twin-Titan RTX for 30 days. An 8-layer fully connected mapping network with $\textbf{w}\in \mathbb{R}^{512}$, initial learning rate 0.01, and activation fused leaky Rectified Linear Unit (ReLU) is used. The generator $f$ takes constant input $b\in\mathbb{R}^{4x4x64}$, both the generator and discriminator networks use Adam optimizer ($\beta_1=0, \beta_2=0.99)$ and initial learning rate 0.002. The batch size was fixed to 8 for the entire training. Pre-processing includes image cropping to eliminate metadata from image boundaries. Similar to \cite{Karras2019stylegan2} we use logistic loss with R1 regularizer and path length regularizer.  All other hyperparameter settings were as in \cite{Karras2019stylegan2}. The best checkpoint was selected based on the Fréchet Inception Distance (FID) \cite{heusel2017FID} between distributions of 200 real and generated images with the truncation set to 1.


\begin{figure}[!htbp]
\centering
\includegraphics[width=5.2in, height=3.3in]{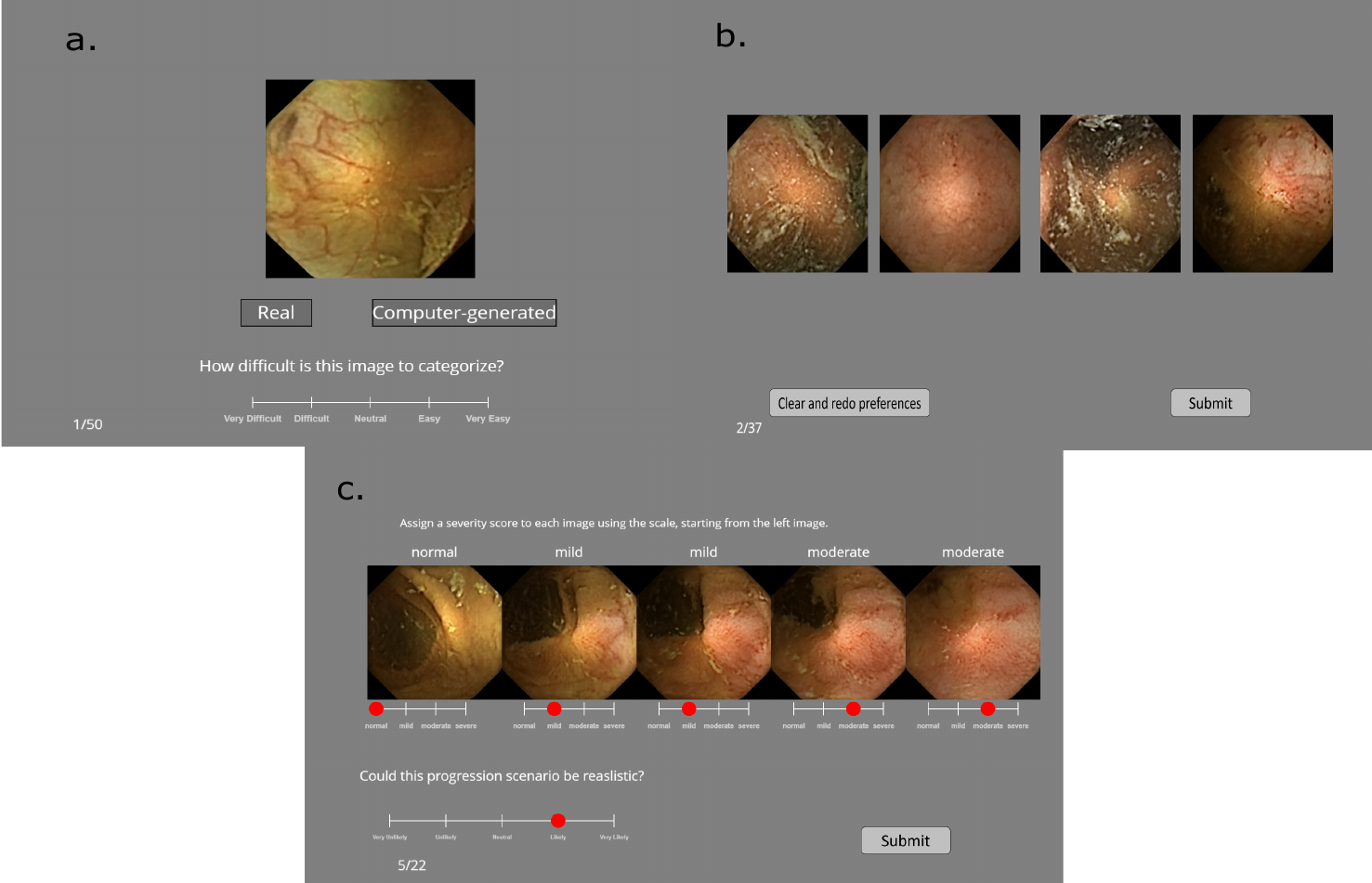}
\caption{The figure illustrates the setup used for subjective online evaluation. (a.) Screen as observed during the visual Turing test for each image. (b.) Screen as observed in the ranking realism task. (c.) Screen as seen in synthetic disease progression task. Please zoom in for clarity, best viewed in color.}
\label{screens}
\end{figure}
\subsection{Task 1: Visual Turing Test} A total of eight experts from Norway, India, and England in the field of gastroenterology performed the Visual Turing test (Fig.\ref{screens} (a)), whereby they were asked to look at an image and indicate whether the image was real or generated in their opinion. With each image, they were also asked to rate the level of difficulty associated with making this choice. For this task, a five-point Likert scale as shown in (Fig.\ref{screens} (a)) with levels between \say{1: very difficult} and \say{5: very easy} was used. The users were given a choice in the number of images they wish to evaluate with each user evaluating at least 50 images. Two users (experience 36 and 28 years respectively) rated 262 images, one user rated 150 images (12 years experience), four users rated 100 images (between 3 and 22 years experience) and one rated 50 images (15 years experience). 

\subsection{Task 2: Ranking realism} Unlike the previous task in which a user rates if a given image is real or generated in isolation, in this task users are shown a set of four images (37 such sets in total) and asked to rank them in order of decreasing realness, the first preference being the most realistic. The users are not told how many in each set are reals or generates, however, each set comprises of two real and two generated images arranged in a random order. Such ranking allows measuring if the generates are plausible enough to be picked over a real one. Six users performed this task.

\subsection{Task 3: Synthetic Disease Progression} In this task, users see a sequence of images that simulate a progression (increase in severity in one way or another) of some disease/abnormality. The task is to assign a severity score to each generate in the sequence as they would in a normal clinical scenario. Fig. \ref{screens}(c) shows this task. Users can assign the same score to more than one image in a sequence. The levels on the scale \say{normal}, \say{mild}, \say{moderate} and \say{severe} are general enough to translate to degrees of severity for different pathologies. The users rate each progression in its realism quality on a five-point Likert scale that goes from \say{very unlikely} to \say{very likely}. Users see 22 such progressions. Six users performed this task.

\section{Results and Analysis :}
\label{sec:analyses}
\textbf{Experimental Setup :} The Graphical User Interface (GUI) for each experiment was designed using the Psychopy experiment builder and made online via the Pavlovia website. A post-experiment survey was attached to the experiment to collect additional data such as years of experience, familiarity with WCE as well as comments on different aspects of the experiment. The GUIs were created similar to the viewing conditions of the Rapid Reader software, which is a commercial software to perform WCE diagnosis. Additionally, the interface and other aspects of the experiment were verified before launch by a gastroenterologist to ensure that the setup was easy to navigate and conduct diagnoses on by doctors. The users had complete control over the time duration for which each image/sequence was looked at, across all experiments. The experiments were designed to be compatible for any screen resolution, however, experts were advised to view it on the largest screen available to them (laptops or other desktop devices) for optimum viewing conditions.

\noindent\textbf{Dataset Details :}  The training dataset comprises of approximately 200k unlabeled images from different capsule modalities. Most of these images (80k) are from the PS-DeVCEM dataset \cite{mohammed2020MARKDATA, AnujaNature} taken with the PillCam Colon 2 Capsule while the other 3478 images taken with the capsule modality Olympus EC-S10 are selected from the abnormal categories in the OSF-Kvasir-Capsule dataset \cite{smedsrud2021kvasir}. The remaining images are from WCE examinations of 10 patients with varying UC activity as well as other pathologies also with the PillCam Colon 2 Capsule, that were conducted at the Gjøvik hospital in Norway in 2021. The images exhibit varying degree of bowel cleanliness.

\begin{table}[h]
\begin{minipage}[b]{0.56\linewidth}
\centering
\begin{tabular}{p{6.665em}cccc}
    \hline
    \multicolumn{1}{c}{UserID} & Images & probability &  95\% CI & p-value \bigstrut\\
    \hline
    \multicolumn{1}{c}{1} & 262 & 0.51 &  0.45 to 0.57 & 0.66 \bigstrut\\
    
    \multicolumn{1}{c}{2} & 262 & 0.58 &  0.52 to 0.64 & 0.009 \bigstrut\\
    
    \multicolumn{1}{c}{3} & 100 & 0.43 &  0.33 to 0.53 & 0.19 \bigstrut\\
    
    \multicolumn{1}{c}{4} & 100 & 0.49 &  0.39 to 0.59 &  0.92 \bigstrut\\
   
    \multicolumn{1}{c}{5} & 100 & 0.54 &  0.44 to 0.64 &  0.48 \bigstrut\\
   
     \multicolumn{1}{c}{6} & 100 & 0.57 &  0.46 to 0.67 &  0.19 \bigstrut\\
   
     \multicolumn{1}{c}{7} & 150 & 0.55 &  0.47 to 0.63 & 0.22 \bigstrut\\
    
     \multicolumn{1}{c}{8} & 50 & 0.42 &  0.28 to 0.56 &  0.32 \bigstrut\\
    \hline
     \multicolumn{1}{c}{Average} & 1024 & 0.52 &  0.49 to 0.55 &  0.07 \bigstrut\\
     \hline
      \end{tabular}
    \caption{Table shows probabilities of identifying generates along with 95\% confidence intervals and p-values for eight users.}
    \label{table:tab1}
\end{minipage}\hfill
\begin{minipage}[b]{0.44\linewidth}
\centering
\includegraphics[width=1.0\linewidth]{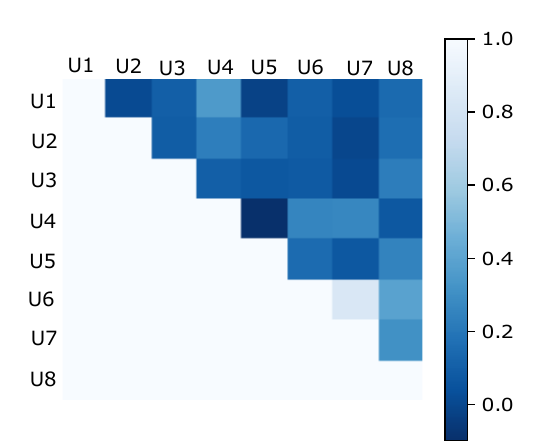}
\captionof{figure}{Krippendorff agreement coefficients among eight users for visual Turing test. Only the upper triangular portion of this symmetric matrix is presented for clarity.}
\label{fig:krpendorf}
\end{minipage}
\end{table}

\noindent\textbf{Task 1: Visual Turing Test}\\ We compared the probability of detecting generates against the probability of random guessing (50\%) for each user and collectively using a two-tailed One-Proportion Z-Test at significance level $\alpha=0.05$. The assumption being tested is that if the generates are as good as real, then doctors can guess correctly only 50\% of the time. As shown in Table \ref{table:tab1} for seven out of eight users, the probability of detection is not significantly different from random guessing (p>0.05), the overall probability for all users being 52.78\% (95\% confidence interval 0.496 to 0.558, p>0.05), which suggests random guessing. However, for UserID 2 who observed 262 images, the detection probability of 58.2\% with p<0.05 (95\% confidence interval 0.52 to 0.64) indicates a probability of identifying generates beyond that by chance. The exact p-values are shown in Table \ref{table:tab1}.

Beyond this, we calculated the agreement scores using the Krippendorff's coefficients \(\alpha_k\) between different users (refer Fig. \ref{fig:krpendorf}). These agreement scores are calculated over 50 images that are common between all users. The overall agreement between eight users is 0.173. Individual agreement scores vary between \(\alpha_k = -0.098\) to \(\alpha_k = 0.83\) whereby the scores for most users are below 0.4, indicating the absence of strong agreement on which images are real or generated among users. The agreement \(\alpha_k = 0.83\) occurs between users 6 and 7 both of whom are from the same medical institution. Although these users randomly guess (see Table \ref{table:tab1}), we assume that there are similarities in the way images are judged and diagnosed, leading to high agreement.
The full $\alpha_k$-matrix of the coefficients is shown in Fig. \ref{fig:krpendorf}.

\begin{figure}[!htbp]
\centering
\includegraphics[width=\linewidth, height=2in]{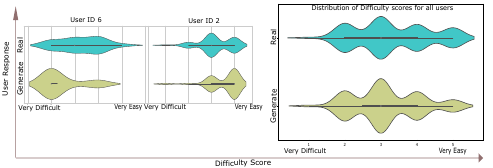}
\caption{Figure shows the distribution of difficulty/easiness of the Turing test as rated by users. Although individual opinions on the difficulty level of images vary between users (example UserID 2 and 6), the overall distribution of the score averages out with a slightly longer tail towards easy than difficult.}
\label{violins}
\end{figure}

In Fig. \ref{violins} we compare the difficulty score ratings of UserID 2 (not guessing randomly) with another user (UserID 6 who guesses randomly). UserID 2 considers the Turing test to be \say{easy} on average when identifying images and is correct 58\% of the times, on the other hand UserID 6 who is correct 57\% of the times considers the choice of identifying a generate more difficult. The distribution of difficulty scores as rated by all users during the test is shown in Fig. \ref{violins}.

\begin{figure}[!htbp]
\centering
\includegraphics[width=\linewidth]{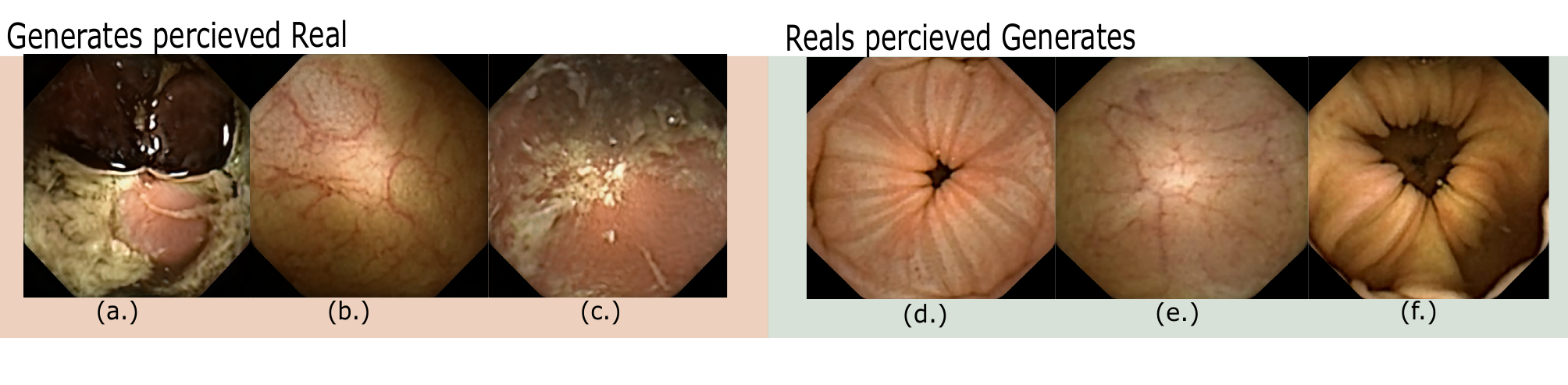}
\caption{Figure illustrates images that were labeled wrongly by the experts with overwhelming majority. All experts labeled image (a.) as real and 7/8 experts labeled images (b.) and (c.) as real while they are actually generated. Similarly, all experts labeled image (d.) to be a generate and 7/8 experts labeled images (e.) and (f.) to be generates
while they are actually real.}
\label{spec_ims}
\end{figure}

In Fig. \ref{spec_ims} we show interesting images from the Turing test where experts agree on an incorrect response in overwhelming majority. In other words, these are the cases where a generated image is identified as real by a majority of the experts and vice versa.

\begin{figure}[!htbp]
\centering
\includegraphics[width=\linewidth]{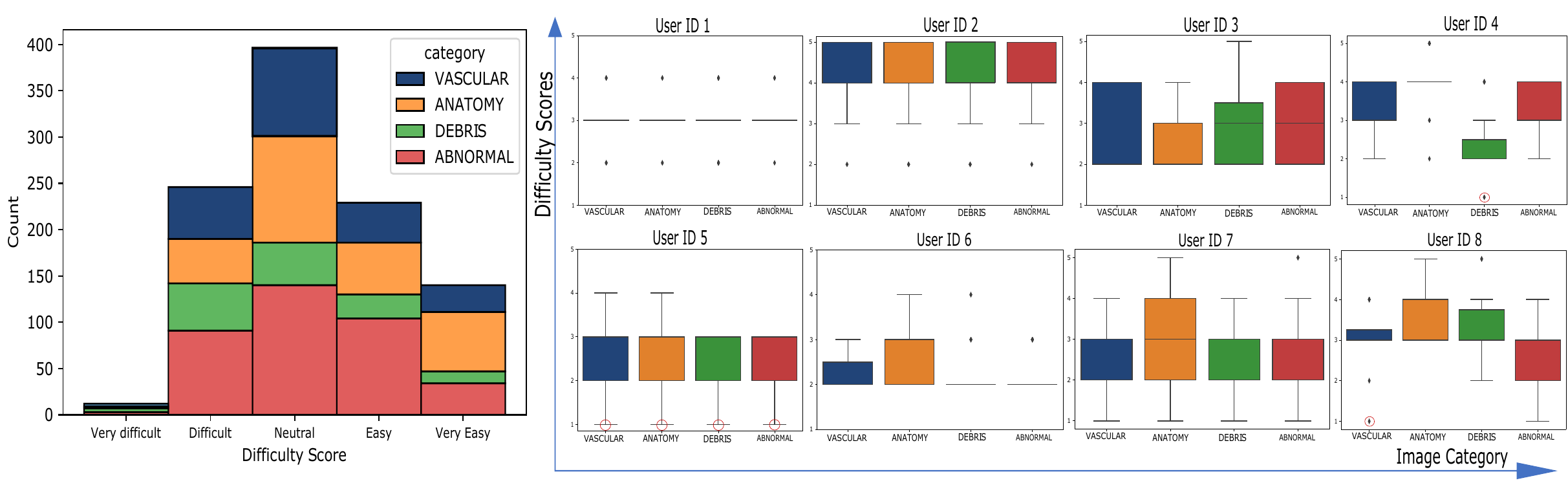}
\caption{The figure shows the overall (left histogram) and individual (right boxplots) difficulty ratings across four generate categories. The red circles indicate selected outliers rated 'Very difficult' and are analyzed in Fig. \ref{outliers}.}
\label{category}
\end{figure}

\begin{figure}[!htbp]
\centering
\includegraphics[width=0.8\linewidth]{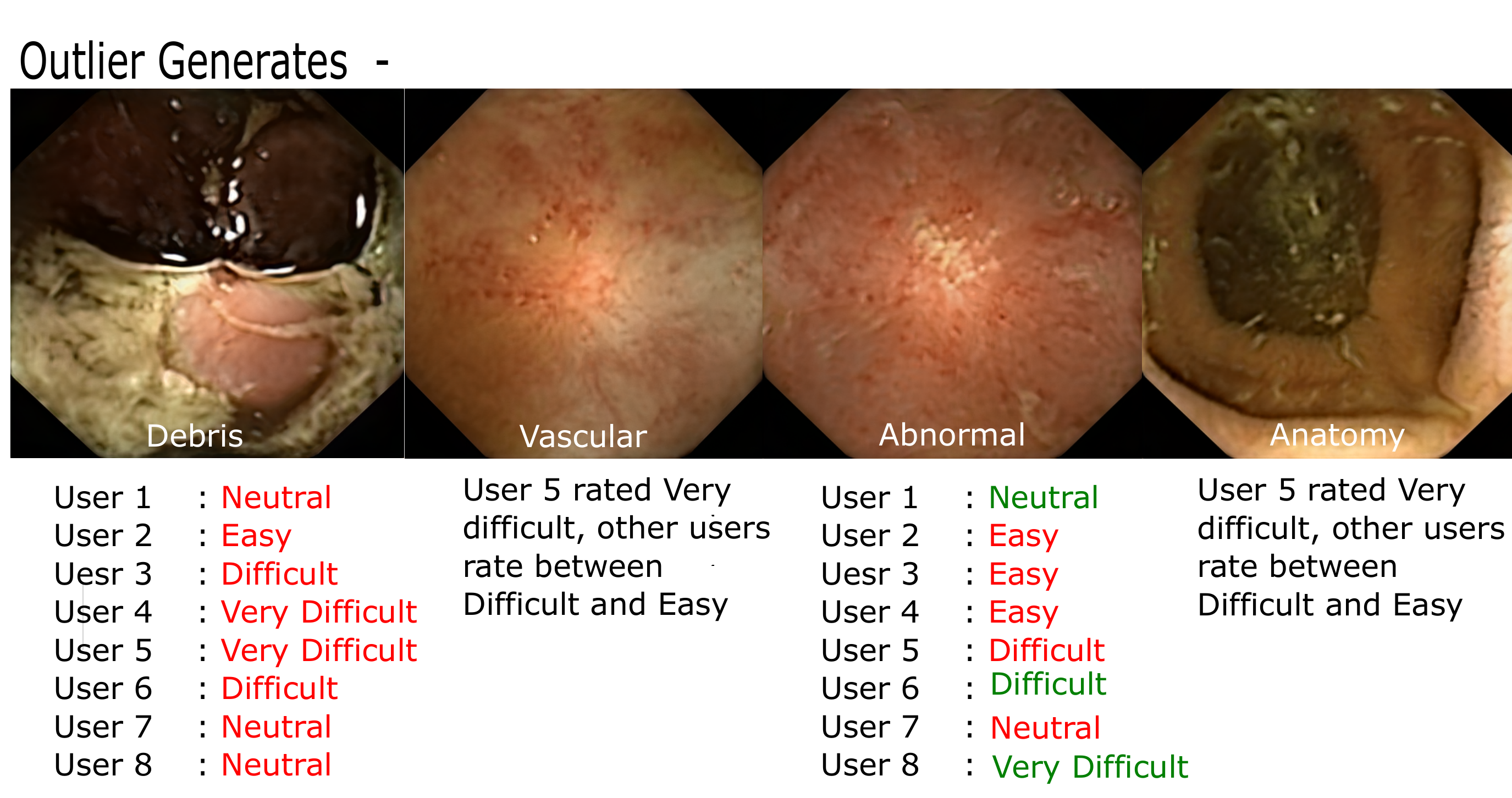}
\caption{The figure shows some outlier generates that one or more experts perceive very difficult to classify. The rating in red text signify when user is wrong whereas green signifies when a user identified the generate correctly. It is interesting to see that (a) all users thought image 1 was real when it was generated and (b) users who perceived image 3 (abnormal) to be easy were incorrect.}
\label{outliers}
\end{figure}

Furthermore, we assign all generated images to one of four general categories given by \say{vascular}, \say{anatomical}, \say{debris}, \say{abnormal}. These categories are based on the visual changes that occur along the attributes $a =[a_1, a_2, ... a_n]$ from which an image comes. Vascular refers to changes in vascular patterns within images (clear as opposed to blurry, refer to images (b.) and (e.) in Fig. \ref{spec_ims}), which are subtle but strong indications of degrading health. Anatomical refers to the category by which images undergo anatomical transformation. These include changes in mucosa (mucosa of different section of intestines is markedly different and can be identified by the muscosal surface), capsule orientations (facing lumen view, facing mucosa view, or changing from one view to another along the attribute) as well as any other anatomical changes largely of normal nature. The debris category is assigned to those images that simulate gastrointestinal content/debris in images (refer to images (a.) and (c.) in Fig. \ref{spec_ims}). Lastly, category abnormal is reserved for images that simulate different abnormal/pathological conditions. In Fig. \ref{category}, we analyze if categories affect the difficulty rating while performing the Turing test, in other words, are generates from some categories more difficult to classify? From the histogram in Fig. \ref{category}, perceived difficulty is seen to be relatively independent of the category, with the overall perceived difficulty varying similarly for all categories. Individual difficulty ratings indicate the presence of outliers (red circles in Fig. \ref{category}). We show some of these in Fig. \ref{outliers} where at least one of the experts rates \say{Very Difficult}. 

\noindent\textbf{Task 2: Ranking Realism}\\ As indicated above, the aim of this task is to quantitatively measure if the generates can be perceived to be visually better (more realistic) than real images, when looked at simultaneously. The reals and generates that comprise the quadruplets are chosen randomly from the real and generated databases. Table \ref{tab:exp2} shows the percentage by which the generates were ranked higher than real WCE images on realness. On average, users perceive a generate to be the most realistic (given first preference, even over reals) 63.5\% of the time, while both generates are ranked higher than reals 10.8\% of the time.

\begin{table}[!ht]
\renewcommand*{\arraystretch}{1}
 \centering
    \begin{tabular}{p{2em}ccc}
 \multicolumn{1}{c}{UserID} & \% \(1^{st}\) preference generate &  \% Both preferences generates   \bigstrut\\
    \hline
    \multicolumn{1}{c}{1} & 56.75\% & 18.91\%  \bigstrut\\
 
    \multicolumn{1}{c}{2} & 29.72\% &  5.41\%  \bigstrut\\
    \multicolumn{1}{c}{4} & 81.08\% & 21.62\%  \bigstrut\\
    \multicolumn{1}{c}{5} & 81.08\%  & 5.41\%   \bigstrut\\
    \multicolumn{1}{c}{6} &  67.56\% & 10.81\% \bigstrut\\
    \multicolumn{1}{c}{7} & 64.86\% &  2.70\% \bigstrut\\

    \hline
        \end{tabular}
   \caption{ The first column shows the \% of times a generate was ranked first (most realistic) among four choices, two of which were real. The second column shows as percentage, the number of times both the generates were chosen to be more realistic than reals out of 37 trials.}
  \label{tab:exp2}
\end{table}

\begin{figure}[!ht]
\centering
\includegraphics[width=5.5in,height=5.5in]{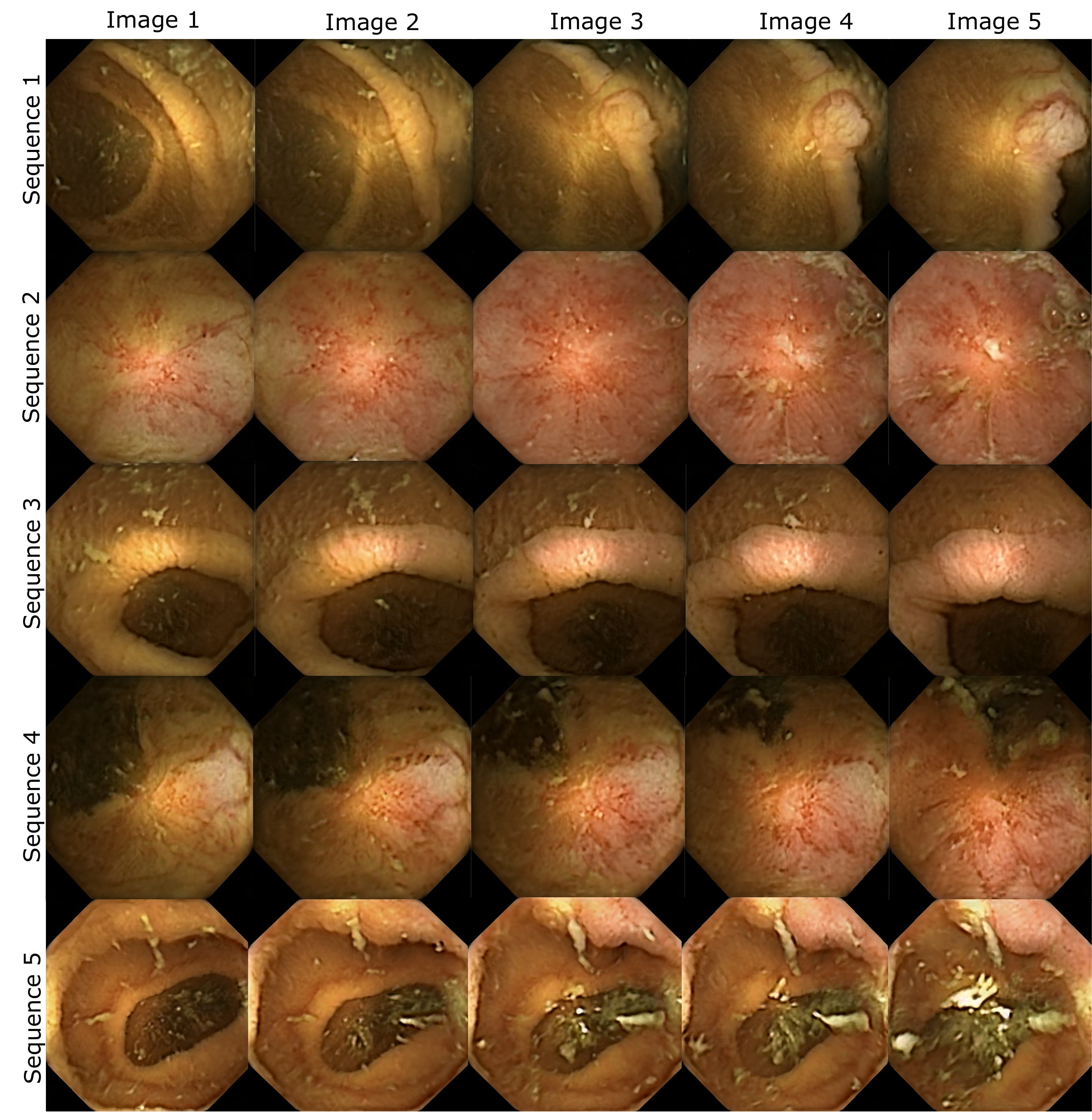}
\caption{The figure shows five disease progressions shown to the experts as part of task 3. The severity increase from left (image 1) to right (image 5). The severity ratings and plausibility histogram corresponding to each sequence is shown in Fig.\ref{progression2} below. }
\label{progression}
\end{figure}

\begin{figure}[!ht]
\centering
\includegraphics[width=\linewidth]{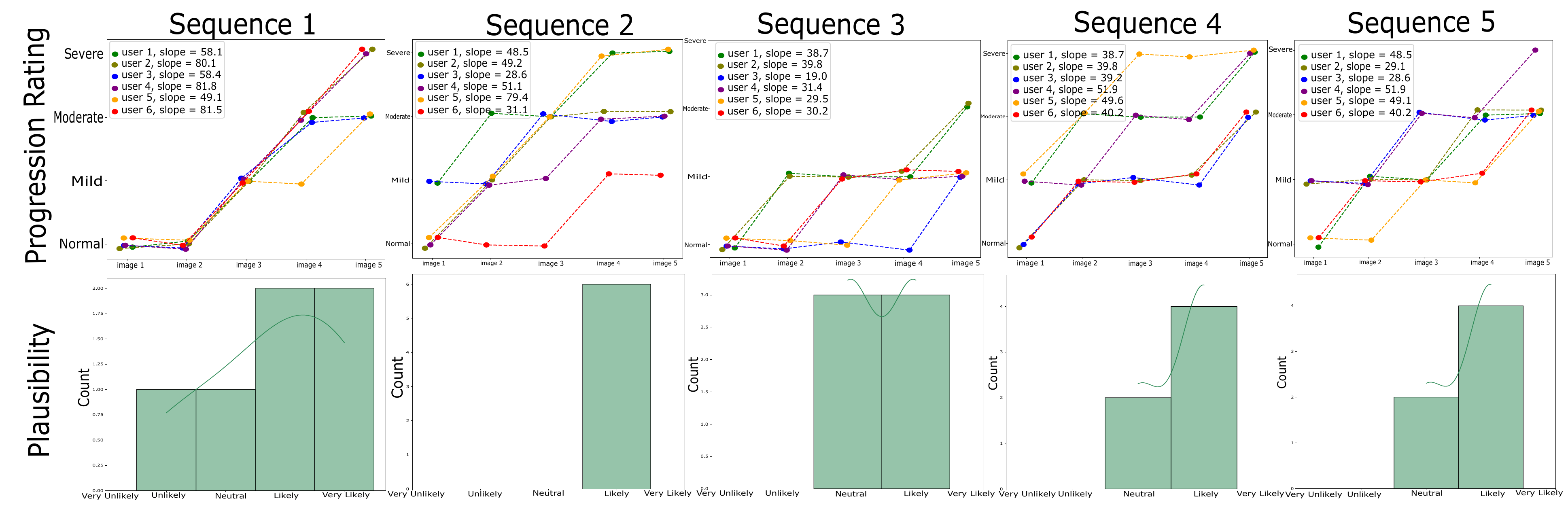}
\caption{The figure shows progression ratings and plausibility histograms for progression sequences illustrated in Fig. \ref{progression}. The progression rating graphs (top row) indicate the severity ratings (y-axis) corresponding to each image (x-axis). The y-axis labels are Normal, mild, moderate and severe (enlarged in the first plot and are same for all plots). The slope of each line is indicated in the legend (please zoom in for clarity. The plausibility curves (bottom row) show a histogram of plausibility rating of each progression by all experts. The x-axis ratings are: Very Unlikely, Unlikely, Neutral, Likely, Very Likely.}
\label{progression2}
\end{figure}

\noindent\textbf{Task 3: Synthetic disease progression}\\ In this task, we measure the quality of disease progressions generated by interpolation along clinically relevant attributes, in the latent space of StyleGAN. Not only are the progressions required to be realistic enough, but also monotonic in nature such that an increase in severity of pathological signs and symptoms can be clearly perceived. We test this by allowing users to rate the severity of each image in a synthetic progression and then test for monotonicity. If the progressions are satisfactory in the aspects mentioned above, the perceived severity by experts would increase monotonically along the direction of progression (from normal to mild to moderate and then severe). We show five images to users that are rated on the four-point severity scale. Monotonicity is measured by calculating the average slope along the line that is best fit for severity scores. Due to a 4-point scale on five images, it is expected that two images along the way will be rated the same ($slope=0$ between two points), however, the higher the slope the higher the monotonicity, and vice versa. Along with severity, each user rates the quality of the progression. This is done by indicating a score to the questions \say{Could this progression scenario be realistic?}, that is, how likely are they to encounter such a progression in their clinical practice. We use these scores to rate the quality of each progression. Fig. \ref{progression} shows progressions from task 3. The accompanying progression ratings, slopes along each best-fit line as well as plausibility distribution of each progression on a 5-point scale from very unlikely (1) to very likely (5) is shown in Fig \ref{progression2}. The average plausibility score for 22 progression scenarios as rated by six experts was 3.68 (between Neutral and Likely) (refer Table. \ref{tab:plausibility}). Two users (experience 36 and 22 years) rated all 22 scenarios likely or above. It should be noted that none of the participants found any of the progressions to be \say{Very Unlikely}, while the progressions were rated \say{Very Likely} 14 times by different experts. Three out of six users never rated a progression below Neutral (rating 3). From our post-experiment survey it was found that inflammation, erythema, ulceration, and mucosal patterns in generated images were found to be particularly realistic by experts (as seen in endoscopy).

\begin{figure}[!ht]
\centering
\includegraphics[width=\linewidth]{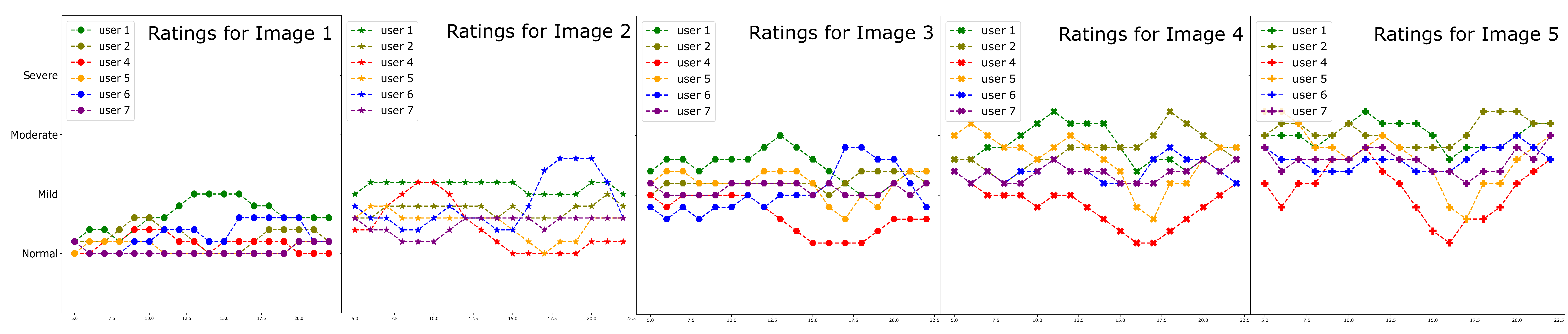}
\caption{The plots show rating patterns of all users corresponding to images with different severities in the progression sequences. Plot 1 shows how each user rated the first image, while  plot 5 shows how users rated the final image (most severe) across 22 sequences. The plots indicate the rolling mean values with window size 5.}
\label{scales}
\end{figure}

Furthermore, progressions provide a tool to measure the subjectivity in how different doctors assess the severity of the same condition on a medical scale. The same image can be perceived as mild inflammation (score 2, no intervention required) by one doctor, while another perceives it as a moderate case (severity 3, which requires treatment and medication). We observed similar subjectivity between the ratings of different doctors in task 3. Fig. \ref{scales} shows the rolling means for severity ratings by different users corresponding to each image in progression sequences. Plot 1 shows the ratings for \say{image-1} of 22 sequences by six users, while plot 5 shows the ratings for the last image. As expected, the plots exhibit an upward trend (lower to higher rating ) from image 1 to image 5. Moreover, the curves clearly expose subjectivity related to the use of the clinical scale among different experts. UserID 1 (experience 28 years) rates higher than other experts on the same image, while userID 4 (experience 5 years) on an average rates lower than other experts. 

\begin{table}[!htbp]

\renewcommand*{\arraystretch}{1}
 \centering
    \begin{tabular}{p{2em}ccccc}
    \hline
    \multicolumn{1}{c}{UserID} &  Average slope & Plausibility score &  Years of experience &  WCE expertise  \bigstrut\\
    \hline
    \multicolumn{1}{c}{1} & 36.36 & 3.18 / 5 &  28 &  expert  \bigstrut\\
 
    \multicolumn{1}{c}{2} & 46.36 &4.13 / 5  &  36  &  very familiar \bigstrut\\
    
    \multicolumn{1}{c}{4} & 25.91 &3.77 / 5 &  5 & somewhat familiar \bigstrut\\
    
    \multicolumn{1}{c}{5} & 43.63 &3.77 / 5  &  4 & somewhat familiar \bigstrut\\

    \multicolumn{1}{c}{6} & 32.27 &4.00 / 5 & 22 & expert  \bigstrut\\

     \multicolumn{1}{c}{7} & 40.90 &3.22 / 5 &  12 & expert \bigstrut\\
    \hline
    \multicolumn{1}{c}{Average} & 37.57 & 3.68 / 5 &  -- & -- \bigstrut\\
    \hline
        \end{tabular}
   \caption{The table shows the average slope and plausibility scores per user averaged over 22 progression sequences where the scores are as follows- 1:very unlikely, 2: unlikely, 3:neutral, 4:likely, 5:very likely. It is notable that most expert users (experience 12-36) found the progressions to be on average likely and their ratings concur with the assumed monotonicity in severity.}
  \label{tab:plausibility}
\end{table}

\section{Applications}
This work is multidisciplinary and therefore finds application in more than one area. In this section we would like to outline potential uses of this work. Firstly, the dataset of synthetic high-quality WCE images can be utilized directly for a number of supervised and unsupervised deep learning algorithms such as abnormality detection and classification, pathology segmentation, organ classification, domain adaptation from one capsule modality to another, etc. Since synthetic data does not have the same privacy and security constraints as the original medical data, its use and distribution is easier. Secondly, this work allows the creation of a high quality synthetic atlas (first to the best of our knowledge) for WCE. The atlas comprises not only of diverse abnormality scenarios but also realistic disease progressions from uncorrelated patients. This is helpful for the training younger gastroenterologists in WCE. Lastly, synthetic progressions as created in this work can prove to be a very useful tool for percolating knowledge from the medical field such as how medical experts rate. A deeper analyses of such inter-rater subjectivity can be carried out to understand the reason behind and find ways to reduce such subjectivity. Conversely, understanding how experts rate generates knowledge that can be leveraged back into deep learning algorithms for, say, severity detection.

\section{Conclusion}
We applied StyleGAN2 for the synthetic generation of high-quality and high-resolution WCE images with all the commonly observed variations. These images are available publicly for use both in data augmentation of deep learning models and for training in WCE diagnoses (https://www.ntnu.edu/web/colourlab/software). The quality and usability evaluation of generated images through three subjective experiments performed by a total of eight experts shows that real and generated images are very hard to be distinguished beyond random guessing. Experts found the progressions to be useful and some of them very likely. Furthermore, this research shows how existing studies from uncorrelated patients, medical conditions, and institutions can be used to learn coherent disease patterns. Such disease patterns have ample uses in a number of computer-aided diagnostic and medical education areas. 

\section*{Data availability}
The dataset OSF Kvasir are publicly available (\url{https://osf.io/dv2ag/}) \cite{smedsrud2021kvasir}, information about dataset Ps-DeVCEM \cite{mohammed2020MARKDATA} is available upon request by Anuja Vats (or authors in \cite{mohammed2020MARKDATA}).

\bibliography{sample}

\section*{Acknowledgements}
This work is part of the project titled “Improved Pathology Detection in Wireless Capsule Endoscopy Images through Artificial Intelligence and 3D Reconstruction” and is funded by the Research Council of Norway (Project number: 300031). We would like to acknowledge the contributions of Dr. Øistein Hovde, Dr. Mark McAlindon, Dr. Varun Nagpal, Dr. Sandeep Ratra, Dr. Snorri Olafsson, Dr. Heidi Kvarme, Dr. Sime Vatn and Dr. Melissa Hale whose involvement in the study has helped strengthen our findings. The authors are extremely grateful for their time and effort in this work.

\section*{Author contributions statement}

A.V, M.P, A.M conceived the study, A.V designed the online experiments with suggestions from  M.P, A.M, Ø.H. The results were analysed by A.V, M.P, A.M. A.V. wrote the manuscript, and all authors reviewed and revised the manuscript. 

\section*{Competing interests}
The authors declare no competing interests.

\end{document}